\documentclass[11pt,twoside]{article} 
\usepackage{asp2004}
\usepackage{epsf}
\usepackage{psfig}

\begin{document} 
\ifx\href\undefined\else\hypersetup{linktocpage=true}\fi

\title{The Crystal Method: Asteroseismology of BPM 37093}

\author{T.S. Metcalfe,$^1$ M.H. Montgomery,$^2$ and A. Kanaan$^3$} 

\affil{$^1$Harvard-Smithsonian Center for Astrophysics\\
$^2$Department of Astronomy, University of Texas-Austin\\
$^3$Departamento de F{\'\i}sica, Universidade Federal de Santa Catarina}

\begin{abstract} 
More than 40 years have passed since Ed Salpeter and others predicted that
the carbon/oxygen cores of the coolest white dwarf stars in our Galaxy
will theoretically crystallize. This effect has a dramatic impact on the
calculated ages of cool white dwarfs, but until recently we have had no
way of testing the theory. In 1992, pulsations were discovered in the
massive potentially crystallized white dwarf BPM~37093, and in 1999 the
theoretical effects of crystallization on the pulsation modes were
determined. Observations from two Whole Earth Telescope campaigns in 1998
and 1999, combined with a new model-fitting method using a genetic
algorithm, are now giving us the first glimpse inside of a crystallized
star.
\end{abstract}

\section{Crisis in the Cosmos}

In 1994, Ed Nather was reading the newspaper and came across a headline
which read, {\it ``Crisis in the Cosmos - stars older than the
universe''}.  Reading through the accompanying article, he learned that a
group of astronomers had fit cosmological models to new observations from
the refurbished Hubble Space Telescope and had concluded that the Universe
was between 8 and 12 billion years old. Meanwhile, another group of
astronomers had used stellar models to fit main-sequence isochrones to
observations of a globular cluster and had derived an age of about 18
billion years. When presented with these details, Ed's reaction was:
``Where's the crisis? Either the cosmological models are wrong, the
stellar models are wrong, or both!'' As we now know, the age estimates
from these two methods later met somewhere in between.

The lesson here is that it is always useful to have independent methods
of measuring a quantity, because it gives us a chance to improve our 
models. In the context of asteroseismology, if we restrict the range of 
our search for pulsation models to be consistent with the spectroscopic 
determinations of the surface gravity and effective temperature, we will 
never find a disagreement between the models. A global search allows the 
pulsation periods to speak for themselves, and lends more credibility to 
the final results.

\begin{figure}[ht]
\epsfxsize=3.0in
\centerline{\epsffile{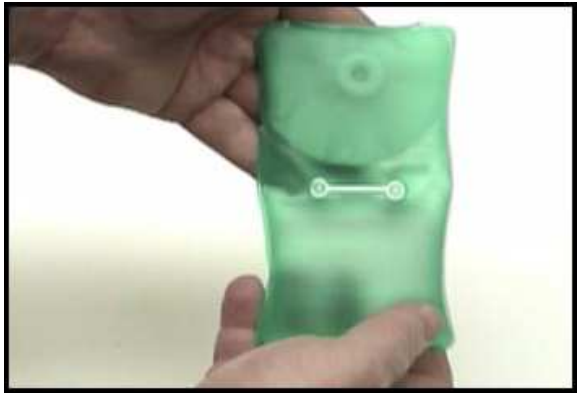}}
\caption{A super-saturated solution of sodium-acetate offers a practical 
example of the latent heat release associated with crystallization.
\label{fig1}}
\end{figure}

\section{Theory}

There is a very simple explanation for how crystallization affects the
derived ages of cool white dwarfs: the process releases latent heat. This
is familiar to anyone who has ever used a small packet of super-saturated
sodium-acetate solution (see Fig.~\ref{fig1}) to keep their hands
warm---the transition from liquid to solid changes the entropy of the
substance, and the difference is released as thermal energy. In a white
dwarf, this new source of thermal energy causes a delay in the gradual
cooling of the star \citep[e.g., see][]{fbb01}.

In addition to the latent heat from crystallization, there is another
source of energy that can delay the cooling even further. When a mixture
of carbon and oxygen crystallizes, the two elements are expected to make
the transition from liquid to solid at slightly different rates
\citep{lv75}. So, the concentration of oxygen in the resulting
solid will be greater than in the liquid \citep{sc93}. This leads to a net
redistribution of oxygen inward and carbon outward (phase separation)
during the crystallization, releasing gravitational potential energy as
additional heat \citep{seg94,sal97,mon99}.

Together, crystallization and phase separation produce a total delay 
of 2-3 Gyr in white dwarf cooling. So if we want to use cool white
dwarfs to date stellar populations, we need to model these processes
accurately. Ideally, we could use observations of pulsating white 
dwarfs to probe the interiors and determine the size of the 
crystallized core empirically---allowing us to calibrate the models.

\section{Observations}

The trouble is, typical white dwarfs with masses near 0.6 $M_\odot$
don't theoretically begin to crystallize until they cool down to about
6000-8000 K (depending on their core composition), and this is well below
the temperatures where they are observed to pulsate. More massive white
dwarfs have higher internal pressures, so they can begin to crystallize at
higher temperatures. Prior to the Sloan Digital Sky Survey data, only one
pulsating white dwarf was known that was theoretically massive enough to
be at least partly crystallized: BPM~37093.

\begin{figure}[ht]
\plotone{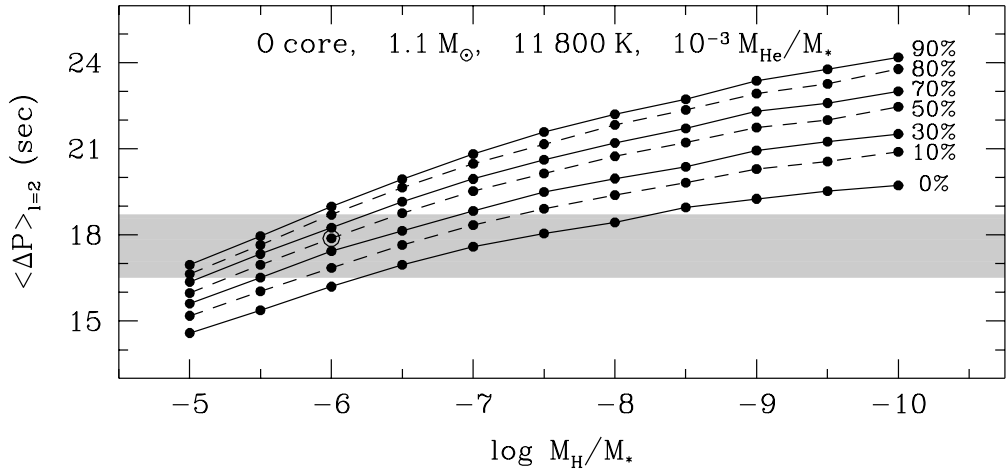}
\caption{Changes to the crystallized mass fraction and the thickness of
the hydrogen layer have similar effects on the mean period spacing in 
models. This degeneracy can be lifted by matching the {\it individual} 
periods.}
\end{figure}

BPM~37093 was discovered to be pulsating by \cite{kan92}. The peak to peak
variation in total light is about 1 percent on a timescale of about 600
seconds, but we also see the signature of beating between closely-spaced
pulsation modes. To resolve the pulsation modes from each other
unambiguously we need to observe the star continuously for a week or more.
So a multi-site campaign of the Whole Earth Telescope \citep{nat90} was
necessary, and in fact there have been two such campaigns---one in 1998
and another in 1999 \citep{kan00,kan04}. In the Fourier Transforms of the
long light curves from these campaigns, we clearly resolve a total of
about 8 distinct pulsation modes, which is what we are attempting to fit
with our theoretical models.

\section{Model Fitting}

In chemically uniform white dwarf models, the pulsation periods are almost
evenly spaced. But abrupt changes in the interior composition cause large
spikes in the buoyancy frequency which can selectively shift {\it some} of
the periods from this simple pattern (mode trapping). The effect of
crystallization is distinct because it alters the pulsation modes by
basically moving the inner boundary from the center of the star out to the
edge of the crystallized core. So, for example, if the star is 50 percent
crystallized, the modes will be confined to the outer half of the 
mass---and this will change {\it all} of the periods.

\subsection{Simple Treatment}

The adjustable parameters in our models include the total mass, the
effective temperature, the masses of the helium and hydrogen layers, the
core composition and the crystallized mass fraction. To make the problem
computationally tractable, early attempts to fit the observations fixed
many of these parameters and tried to match the average spacing between
consecutive $\ell$=2 modes \citep{mw99}. This approach quickly ran into
the difficulty that changes to the crystallized mass fraction and the
thickness of the surface hydrogen layer had very similar effects on the
mean period spacing (see Fig.~2).

\begin{figure}[ht]
\plotone{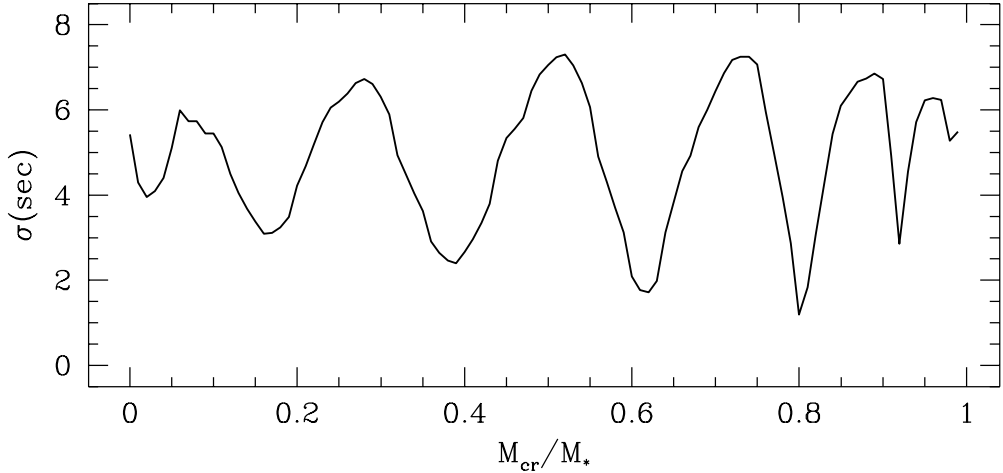}
\caption{Systematic errors in the measurement of $M_{\rm cr}$ can arise 
when fixing some parameters (like the mass, temperature, and core 
composition) incorrectly, allowing secondary minima to become locally 
optimal.}
\end{figure}

To break the degeneracy between these model parameters, we need to use the
{\it individual} periods in addition to the period spacing.  When we do
so, one of the models stands out as the best. But this is just a simple
analysis---since we have fixed so many of the other model parameters, we
have most likely found a locally optimal match to the periods, rather than
the true global solution.

\subsection{Initial Results}

We have recently published our first steps towards a global solution
\citep{mmk04}. To keep the computing time reasonable, we initially did a
series of fits with masses fixed near the spectroscopic values, and we
tried core compositions of pure C and pure O just to test the extreme
limits. Also, for each combination of parameters we examined only 10
values of the crystallized mass fraction, from 0 to 90 percent. All of our
fits reproduced the periods observed in BPM~37093 at the level of about 1
second, and in every case we found solutions with a large crystallized
mass fraction. But it's important to put these numbers into context by
noting the large uncertainties that come about {\it just from the way we
did the fitting}.

We passed artificial data through the same process to estimate the
systematic errors in the crystallized mass fraction that we should expect
from our limited exploration of the models. We found that because of our
low resolution in the crystallized mass fraction, fixing the mass or
composition incorrectly could lead to systematic errors of at least
$\pm$0.2 (on top of the statistical errors of $\pm$0.1). We can understand
this by looking at a model that is 80 percent crystallized, and trying to
match its pulsation periods with models that are anywhere from 0 to 99
percent crystallized (see Fig.~3). What emerges is a series of secondary
minima spaced about 0.2 apart. When we fix one of the other parameters
incorrectly, the minimum at 0.8 can get shallower and one of the nearby 
secondary minima can become the locally optimal ``best fit''.

\begin{figure}[ht]
\plotone{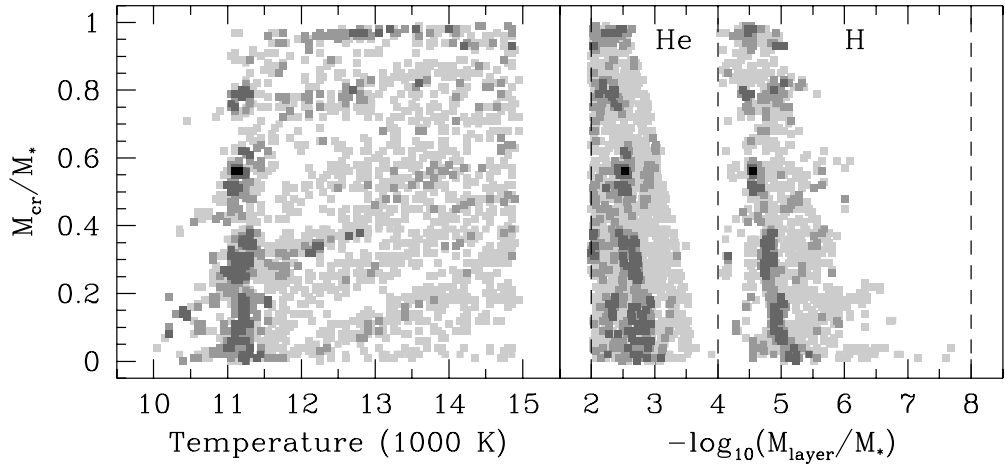}
\caption{Front and side views of the genetic algorithm search space, 
showing better models as progressively darker points. Correlations
between the parameters and multiple secondary minima are evident.}
\end{figure}

So, these initial fitting results tell us two important things: (1)  
BPM~37093 is substantially crystallized, so it's worth spending more
computing time to try to pin down the exact fraction, and (2) we should
probably fit to the nearest 0.01 in the crystallized mass fraction to 
reduce the systematic errors, and we should avoid fixing as many 
parameters as we possibly can.

\subsection{Latest Results}

Our next step has been to treat the crystallized mass fraction as a
completely adjustable parameter to the nearest 0.01. Again, the
computational demands of the problem forced us to fix the mass and the
core composition, but we are performing fits over a broad range of fixed
masses to minimize the systematic errors. The best fit we have found so 
far (fixed $M_*=1.03~M_\odot$, 50:50 C/O core) is significantly better 
than any of the initial fits we published:
\begin{eqnarray*}
T_{\rm eff}          &=& 11,200~{\rm K}             \\
\log(M_{\rm He}/M_*) &=& -2.54                      \\
\log(M_{\rm H}/M_*)  &=& -4.56                      \\
M_{\rm cr}           &=& 0.56~M_*                   \\
\sigma_{\rm P}       &=& 0.52~{\rm sec}             
\end{eqnarray*}
Because of the limited sampling of models with different fixed masses, the
systematic uncertainty on the crystallized mass fraction is still about
$\pm$0.1, but when we have finished we expect to measure this parameter to
within a few percent. There are still secondary minima in the crystallized
mass fraction (see Fig.~4), but we hope to rule these out unambiguously as 
the model-fitting continues. Recently, our mode identification and a 
comparable set of best-fit parameter values were independently found by 
\cite{fb05}.

\section{Summary}

To summarize the main conclusions of this work:

\begin{itemize}

\item Crystallization delays white dwarf cooling by 2-3 Gyr, but we can 
      calibrate the effect on their ages through asteroseismology.

\item BPM~37093 is currently the only pulsating white dwarf massive
      enough to be crystallized, but others are expected from the Sloan
      Digital Sky Survey.

\item Surface layers cause ``mode trapping'', changing the pulsation 
      periods selectively; crystallization squeezes the resonant 
      cavity to change all of the periods.

\item Initial fitting suggests that BPM~37093 is substantially 
      crystallized, and work in progress will measure the crystallized 
      mass fraction to $\pm$3 percent.

\end{itemize}

\acknowledgements{This research was supported in part by the Smithsonian 
Institution through a CfA Postdoctoral Fellowship, and by a small grant 
from NASA administered by the American Astronomical Society.}

\end{document}